\documentclass[twocolumn]{article}

\usepackage[twocolumn,textwidth=18cm,columnsep=.81cm]{geometry}
\usepackage{amsmath}
\usepackage{amssymb}
\usepackage{amsthm}
\usepackage{physics}
\usepackage{graphicx}
\usepackage{hyperref}
\usepackage{dcolumn}
\usepackage{stfloats}
\usepackage{authblk}
\usepackage{helvet}

\usepackage[T1]{fontenc}

\title{A Multi-Class SWAP-Test Classifier}

\author[1]{S. M. Pillay}
\author[2,3]{I. Sinayskiy}
\author[1]{E. Jembere}
\author[2,3,4]{F. Petruccione}
\affil[1]{School of Mathematics, Statistics and Computer Science, University of KwaZulu-Natal, Durban 4001, South Africa}
\affil[2]{School of Chemistry and Physics, University of KwaZulu-Natal, Durban 4001, South Africa}

\affil[3]{National Institute for Theoretical and Computational Sciences (NITheCS), Stellenbosch, South Africa.}
\affil[4]{School of Data Science and Computational Thinking, Stellenbosch University, Stellenbosch 7604, South Africa.}

\date{}

\begin{document}

\twocolumn[
  \begin{@twocolumnfalse}
  \maketitle
    \begin{abstract}
      Multi-class classification problems are fundamental in many varied domains in research and industry. To solve multi-class classification problems, heuristic strategies such as One-vs-One or One-vs-All can be employed. However, these strategies require the number of binary classification models developed to grow with the number of classes. Recent work in quantum machine learning has seen the development of multi-class quantum classifiers that circumvent this growth by learning a mapping between the data and a set of label states. This work presents the first multi-class SWAP-Test classifier inspired by its binary predecessor and the use of label states in recent work. With this classifier, the cost of developing multiple models is avoided. In contrast to previous work, the number of qubits required, the measurement strategy, and the topology of the circuits used is invariant to the number of classes. In addition, unlike other architectures for multi-class quantum classifiers, the state reconstruction of a single qubit yields sufficient information for multi-class classification tasks. Both analytical results and numerical simulations show that this classifier is not only effective when applied to diverse classification problems but also robust to certain conditions of noise.
    \end{abstract}
  \end{@twocolumnfalse}
]

\section{Introduction}
One class of problems that has received considerable attention in quantum machine learning is classification. Quantum kernel methods, that exploit the ability of quantum computers to efficiently perform operations in large Hilbert spaces, have been developed to solve these problems \cite{schuld2019quantum,havlivcek2019supervised,blank2020quantum}. The SWAP-Test classifier is a quantum kernel method that estimates a weighted sum of kernel values for a given test datum and all the training data in parallel \cite{blank2020quantum}. This sum is the result of applying the SWAP-Test, involving only a single qubit measurement, to a quantum state encoding the test datum, training data and their respective labels in a specific format \cite{park2020theory}. In contrast to other quantum kernel methods, the need to prepare the test datum for each estimation of a kernel value is avoided. In addition, a decision rule that makes use of this sum is obtained without the aid of a classical subroutine for training.\\
\indent The SWAP-Test classifier is a binary classifier that can categorise data into only two categories. However, most real-world problems like image recognition, natural language processing and others require multi-class classifiers that can handle more than two classes. To perform multi-class classification with the SWAP-Test Classifier, a number of binary classification models can be used in heuristic strategies like the One-vs-All Strategy \cite{alpaydin2020introduction}. But, these strategies have limitations and the number of binary classifiers required grows with the number of classes. \\
\indent Some of the recent work in quantum machine learning has seen the development of multi-class quantum classifiers that circumvent this growth \cite{nghiem2021unified,perez2020data}. This is achieved by training parametrised circuits, with the aid of classical subroutine, to map the classical data to separate regions in Hilbert space. In some approaches, these regions are centred around label states; a set of quantum states corresponding to the set of classes \cite{nghiem2021unified}. In other approaches, these regions or clusters are constructed, through training, to be as far apart as possible in Hilbert space \cite{perez2020data}. In both these approaches, the choice of label states is inherently tied to the data encoding strategy and vice versa. If, as proposed, orthogonal states in Hilbert space are chosen as the label states then the data encoding strategy will have to utilise a sufficient number of qubits to accommodate the orthogonal states. Alternatively, if many qubits are used for the embedding, then clusters need to be formed around label states in large Hilbert spaces. In this case, the task of finding similarities between the data in these vast spaces may not be easy \cite{huang2021power}. 

This work presents the first multi-class SWAP-Test classifier. Our work is inspired by the binary SWAP-Test classifier and the use of label states in previous work.  With this multi-class quantum classifier, we are able to avoid the cost of constructing multiple binary classifiers to perform multi-class classification.  The use of only single qubit label states, regardless of the data encoding strategy and the number of classes, is novel to this work. Furthermore, the label states are stored separately from the data, ensuring that the choice of label states remains independent from the data encoding strategy. This has the advantage that the number of qubits required and the topology of the circuits used need not change with the number of classes. Importantly, the state reconstruction of only a single qubit is performed regardless of the number of classes or the data encoding strategy. This ensures that the measurement strategy is also independent from the number of classes. In contrast to other architectures for multi-class quantum classifiers, the state reconstruction of a single qubit yields all the necessary information for multi-class classification tasks \cite{zeng2022multi}.

Given some multi-class classification problem, the classifier is realised by first preparing a  quantum state encoding the test datum, the training data and their respective labels in a specific format. The label states are chosen such that their corresponding Bloch vectors, which we refer to as label vectors, are maximally separate on the Bloch sphere. A modified SWAP-Test, involving a state reconstruction of the qubit storing the label states, is performed on the prepared state. This effectively yields a linear combination of label vectors. The contribution of each label vector is a weighted sum of kernel values between the test data and all the training data with that label. The kernel values in this sum are computed in parallel, just as with the binary SWAP-Test classifier. The type of kernel that is evaluated can be tailored to suit the classification problem by changing the data encoding strategy. Finally, the overlap between the resulting vector and each label vector is evaluated classically. The test datum is assigned the label whose label vector achieves the highest overlap. Like the binary SWAP-Test classifier, this multi-class SWAP-Test classifier does not inherently rely on a classical subroutine for training. 

The effectiveness of the multi-class SWAP-Test classifier is demonstrated by applying it to a number of different classification problems, each with a different dimension of features and a different number of classes. The datasets used are generated XOR datasets as well as real-world datasets including Iris (3 classes), Wine (3 classes) and Digits  (10 classes) datasets \cite{Dua:2019}. We show analytically that the multi-class classifier achieves high accuracies on these real-world datasets. Even without training and with only standard data encoding strategies, these experiments show that the multi-class SWAP-Test classifier is remarkably powerful. 

The performance of the multi-class SWAP-Test classifier is then considered under realistic conditions. To do this, we demonstrate the robustness of the classifier to finite sampling and noise. Through variance analysis, we show that the number of label states that can be accurately distinguished on a single qubit grows linearly with the number of repetitions of the required measurements. We also show that, under certain depolarising noise conditions, the classification process remains unaffected. These theoretical results are demonstrated by numerical experiments that incorporate depolarising noise and finite sampling.

The paper is organised as follows: Section 2 outlines the steps that constitute the proposed multi-class swap test classifier, discusses its robustness to noise, and assesses the number of label states that can be stored with a single qubit with this classifier. Section 3 presents the methodology and results of experiments conducted on various datasets. Lastly, Section 4 draws concluding remarks, highlighting possible areas for future work.

\section{Results}
\subsection{Classification with the Multi-Class SWAP-Test Classifier}
\begin{figure*}[htp]
\centering
	\includegraphics[scale=0.3]{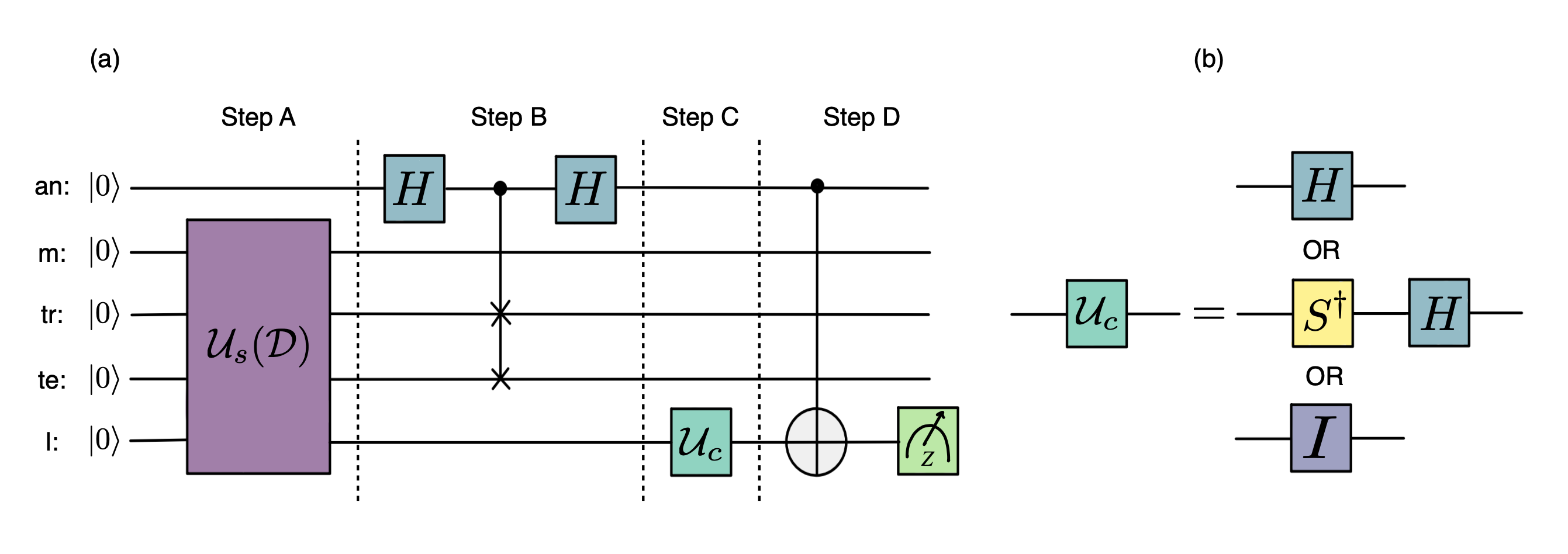}
	\caption{(a) \textbf{The circuit required for the multi-class SWAP-Test classifier.} The first register (an) stores the ancilla. The second register stores the index register (m) which links the training data in the training register (tr) to their respective label states on the label qubit (l). The test data is stored in (te). To perform a state tomography of (l) at the end of the circuit, three circuits performing Steps A and B will be prepared. In each of these circuits, Step A applies $\mathcal{U}_s(\mathcal{D})$ which prepares the test data, training data and training labels in a quantum state $\ket{\Psi_i}$, given in equation (\ref{stateprep}). Step B then swaps the registers containing the test and training data. In each circuit, Step C applies one of the gate sequences in (b) to perform a change of basis to the X-basis, Y-basis or maintain the Z-basis. The three circuits evaluate the predicted vector $\mathbf{y}_{pred}$ which is then used in an assignment function to classify the test data.}
\label{circuit_diagram}
\end{figure*}

Classification is a fundamental problem in machine learning. Given a dataset  

\begin{align}
	\mathcal{D} = \{(\mathbf{x}_i,y_i)\}^{M}_{i=1} \subset \mathbb{R}^N \cross \{y_i\}^{L}_{i=1},
\end{align}

\noindent consisting of training data $\{\mathbf{x}_i\}_{i=1}^M$ and their respective labels $\{y_i\}_{i=1}^M$, the goal of supervised classification is to develop a model for classifying unlabelled data. The algorithms for developing these models are called classifiers. This section describes the steps that constitute the multi-class SWAP-Test classifier. These steps are also outlined in Figure \ref{circuit_diagram}.

For some unlabelled test datum $\mathbf{\tilde{x}}$, the multi-class SWAP-Test classifier requires the test datum, the training data and their respective labels to be encoded in a quantum state as 

\begin{align}
\label{stateprep}
\ket{\Psi_i} = \sum_{m=1}^{M} \sqrt{w_m} \ket{0}  \ket{\mathbf{\tilde{x}}} \ket{\mathbf{x}_m} \ket{y_m} \ket{m}.
\end{align}

\begin{figure*}
\centering
	\includegraphics[width=\textwidth]{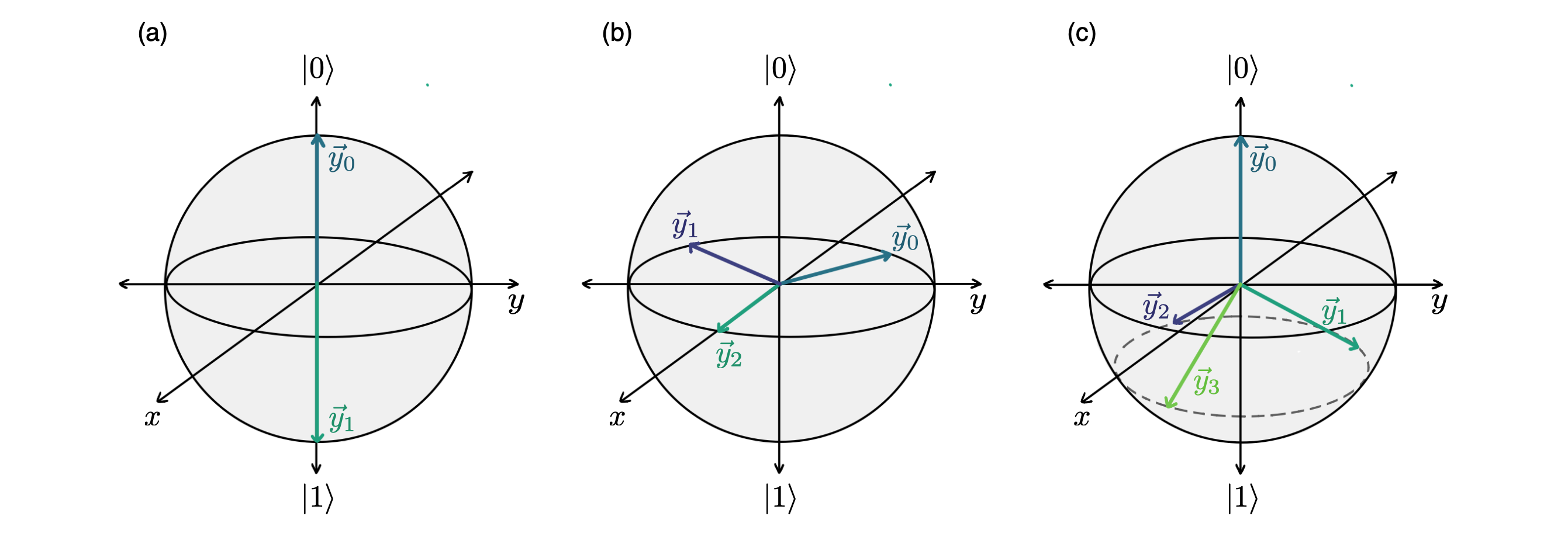}
	\caption{\textbf{The optimal choice of label vectors} for (a) 2 classes $\{y_0:[0,0,1],y_1:[0,0,-1]\}$, (b) 3 classes $\{y_0:[1,0,0],y_1:[-0.5,0.866,0],y_2:[-0.5,-0.866,0]\}$ and (c) 4 classes $\{y_0:[0,0,1],y_1:[-0.471,0.861,-0.333],y_2:[-0.471,-0.861,-0.333],y_2:[0.943,0,-0.333]\}$. These vectors point to solutions of the Tammes problem.}
		\label{label_vectors}
\end{figure*}

\noindent The first qubit is initialised in the ground state $\ket{0}$ and will later be used as the ancilla in a modified SWAP-Test. The index register $\sum_{m=1}^{M} \sqrt{w_m}\ |m\rangle$ with $\sum_{m=1}^{M}w_m=1$ is used to link the training data to their respective labels. The test and training data are encoded in separate registers. Some unitary operator $\mathcal{U}_{\Phi(\mathbf{x})}$ encodes the test and training data into quantum states as $|\mathbf{\tilde{x}}\rangle = \mathcal{U}_{\Phi(\mathbf{\tilde{x}})}|0\rangle$ and $|\mathbf{x}_m\rangle = \mathcal{U}_{\phi(\mathbf{x}_m)}|0\rangle$, respectively. This operation can be understood as applying a feature map to the classical data \cite{schuld2019quantum}. The choice of operator, corresponding to a choice in encoding method, decides the type of kernel that will be evaluated by the classifier \cite{schuld2018supervised,schuld2021supervised}. The label qubit $|y_m\rangle$ is used to store the label state that each label $y_m$ is mapped to. Each label state 

\begin{align}
	|y_m\rangle = 
	\mathrm{cos}\bigg(\frac{\theta_{y_m}}{2}\bigg)\ |0\rangle + 
	\mathrm{e}^{i\phi_{y_m}}\ \mathrm{sin} \bigg(\frac{\theta_{y_m}}{2}\bigg)\ |1\rangle
\end{align}

\noindent with $0\le\theta_{y_m}\le\pi$ and $0\le\phi_{y_m}\le2\pi$ has an associated Bloch vector

\begin{align}
	\mathbf{y}_m = 
	\begin{pmatrix}
	\mathrm{cos}\phi_{y_m}\mathrm{sin}\theta_{y_m}\\
	\mathrm{sin}\phi_{y_m}\mathrm{sin}\theta_{y_m}\\
	\mathrm{cos}\theta_{y_m}
	\end{pmatrix}	
\end{align}

\noindent which will be referred to as the label vector of $y_i$. 
The label states are chosen such that their Bloch vectors are as far apart as possible on the Bloch sphere. For this classifier, we propose that the optimal placement of Bloch vectors be identified in solutions to the Tammes problem: the problem of placing $L$ points on a unit sphere so that the two closest points are as far apart as possible \cite{tammes1930origin}. This allows maximal separation of vectors around the Bloch sphere. Some of these placements, for $L=2,3$ and $4$,  are shown in Figure \ref{label_vectors}.

Following the required state preparation, a modified SWAP-Test is performed on the test and training data registers. A \texttt{C-SWAP} gate conditioned on the ancilla, in between two Hadamard gates applied to the ancilla, swaps the two registers. The result is

\begin{align}
	{\ket{\Psi_f}} & = H_a.\texttt{C-SWAP}.H_a{\ket{\Psi_i}}\nonumber \\
	& = \sum_{m=1}^{M} 
	\frac{\sqrt{w_m}}{2}
	(\ket{0}\ket{\psi_+} + \ket{1}\ket{\psi_-})
	\ket{y_m}
	\ket{m}
\end{align}

\noindent with $\ket{\psi_{\pm}} = \ket{\mathbf{\tilde{x}}}\ket{\mathbf{x}_m} \pm\ket{\mathbf{x}_m}\ket{\mathbf{\tilde{x}}}$. The subscript $a$ denotes that the Hadamard gate $H$ is applied to the ancilla. 

In order to perform the necessary state tomography, the above state needs to be prepared three times. Then, the required change of basis on each of the three states may be performed:

\begin{align}
		\ket{\Psi_{fx}} & =  H_l\ket{\Psi_f},\nonumber\\ 
		\ket{\Psi_{fy}} & =  H_lS^\dagger_l\ket{\Psi_f},\\
		\ket{\Psi_{fz}} & =  \ket{\Psi_f}.\nonumber
\end{align}

\noindent Here, the subscript $l$ denotes that the gates $H$ and $S^\dagger$ are applied to the label qubit. 

Before we perform any measurement on the states, a \texttt{C-NOT} operation controlled on the ancilla and targeted on the label qubit is applied to each state:

	\begin{align}
		\ket{\tilde{\Psi}_{fx}} & =  \texttt{c-not}_{a,l}\ket{\Psi_{fx}},\nonumber\\ 
		\ket{\tilde{\Psi}_{fy}} & =  \texttt{c-not}_{a,l}\ket{\Psi_{fy}},\\
		\ket{\tilde{\Psi}_{fz}} & =  \texttt{c-not}_{a,l}\ket{\Psi_{fz}}.\nonumber
	\end{align}

\noindent This converts what would be a two qubit measurement to a single qubit measurement in each case \cite{park2021robust}. Finally, the measurement of a single qubit observable $\langle \sigma_z^l \rangle $, where the superscript $l$ indicates that the operator acts only on the label qubit, is performed on each state $\rho_{fs}=|\tilde{\Psi}_{fs} \rangle \langle \tilde{\Psi}_{fs}|$ for $s \in \{x,y,z\}$. The results of these measurements is used to construct a vector which we refer to as the predicted vector $\mathbf{y}_{pred}$

\begin{align}
	\mathbf{y}_{pred} &=
	\begin{pmatrix}
	\mathrm{Tr} \big( \sigma_z^{(l)} \rho_{fx} \big)\\
	\mathrm{Tr} \big( \sigma_z^{(l)} \rho_{fy} \big)\\
	\mathrm{Tr} \big( \sigma_z^{(l)} \rho_{fz} \big)
	\end{pmatrix}, \nonumber\\
	\\
	&=\begin{pmatrix}
	\sum_m \mathrm{w}_m | \langle \mathbf{\tilde{x}} |\mathbf{x}_m\rangle|^2 \mathrm{cos}\phi_{y_m}\mathrm{sin}\theta_{y_m}\\
	 \sum_m \mathrm{w}_m | \langle \mathbf{\tilde{x}} |\mathbf{x}_m\rangle|^2 \mathrm{sin}\phi_{y_m}\mathrm{sin}\theta_{y_m}\\
	 \sum_m \mathrm{w}_m | \langle \mathbf{\tilde{x}} |\mathbf{x}_m\rangle|^2 \mathrm{cos}\theta_{y_m} 
	\end{pmatrix}\nonumber.
\end{align}

At first, the significance of $\mathbf{y}_{pred}$ may not seem clear. However, we can use the fact that the fidelities that result from the modified SWAP-Test represent a valid kernel $k(\mathbf{\tilde{x}},\mathbf{x}_m) = |\langle \mathbf{\tilde{x}} |\mathbf{x}_m\rangle|^2$. Then, if $\alpha_i = \sum_{m|\mathrm{y}_m=i} \mathrm{w}_m k(\mathbf{\tilde{x}},\mathbf{x}_m)$ the predicted vector may be expressed as 

\begin{equation}
	\mathbf{y}_{pred} = 
	\sum_{i=1}^{L}
	\alpha_i 
	\begin{pmatrix}
		\mathrm{cos}\phi_{y_i}\mathrm{sin}\theta_{y_i}\\
		\mathrm{sin}\phi_{y_i}\mathrm{sin}\theta_{y_i}\\
		\mathrm{cos}\theta_{y_i} 
	\end{pmatrix}
\end{equation}

\noindent or equivalently, 

\begin{equation}
\label{y_pred_eq}
	\mathbf{y}_{pred} = \sum_{i=1}^{L} \alpha_i \mathbf{y_i}.
\end{equation}

\noindent Now, it is apparent that the predicted vector is a linear combination of the label vectors. The weight of each label vector, $\alpha_i$, is the sum of the kernel values between the test data and the training data that have that label. A high $\alpha_i$ increases the overlap between the $\mathbf{y}_{pred}$ and $\mathbf{y}_i$ and indicates a high similarity between the test datum and the training data belonging to class $y_i$. 

\begin{figure*}[htp]
\centering
	\includegraphics[width=\textwidth]{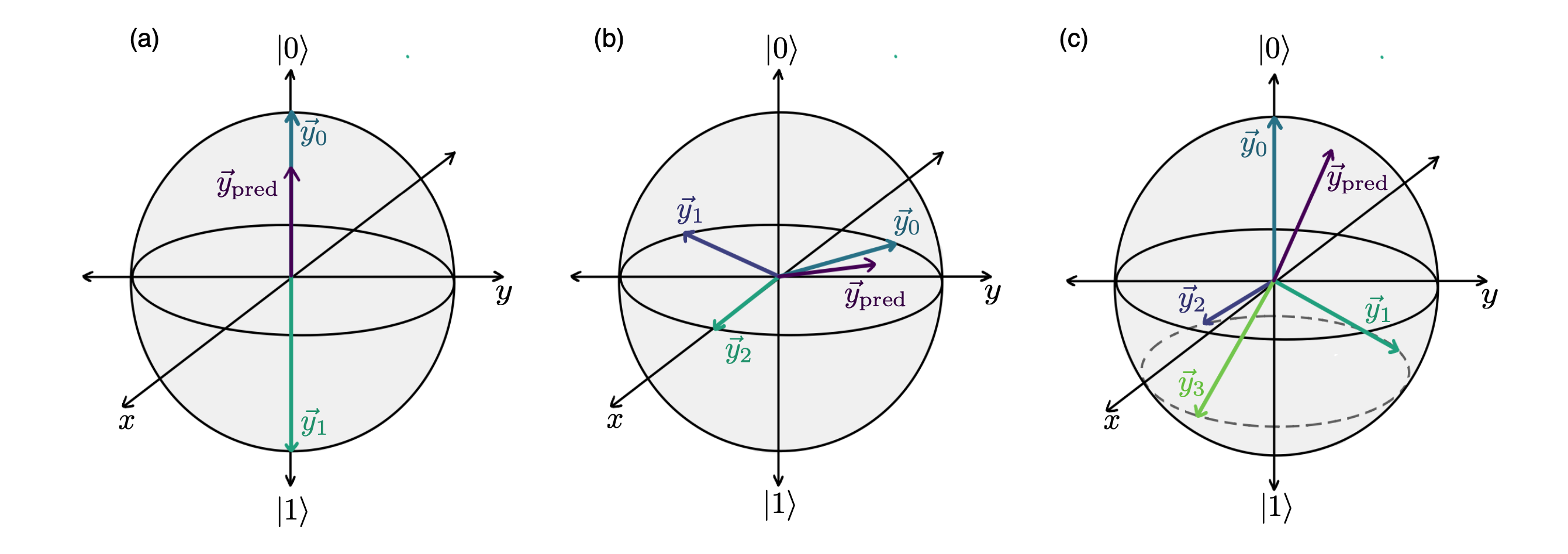}
	\caption{\textbf{Illustrative predicted vectors that could be obtained by this multi-class SWAP-Test classifier for problems with (a) 2 classes, (b) 3 classes and (c) 4 classes}. According to the assignment function given in equation \ref{assignment_function}, the test point will be assigned to the class $y_i$ when the inner product between its label vector $\mathbf{y_i}$ and the obtained predicted vector is highest. According to the definition of the inner product, the inner product will be its highest when the angle between the vectors is the smallest. In each of these diagrams, the test point will be assigned to class $y_0$. \emph{Note}: these predicted vectors are not obtained from any concrete classification problem.}
	\label{predicted_vectors}
\end{figure*}
The predicted vector that has been estimated by the quantum circuits is then used in the following assignment function
\begin{equation}
\label{assignment_function}
	\tilde{y} = \mathrm{max}_{y_i}\{\mathbf{y_i}\cdot \mathbf{y}_{pred}\}.
\end{equation}

\noindent Effectively, the test datum  is assigned the class $y_i$ when the inner product between the label vector $\mathbf{y_i}$ and the predicted vector $\mathbf{y}_{pred}$ is the highest. This indicates that $\mathbf{y_i}$ and $\mathbf{y}_{pred}$ have the highest overlap. Figure \ref{predicted_vectors} shows a few examples of predicted vectors that could be obtained by the classifier and how the test datum that generated the predicted vector would be classified. 

\subsection{Analysis of Multi-Class SWAP-Test Classifier}

Even though great strides have been made in the development of quantum hardware, two sources of error are inevitably encountered when executing quantum algorithms on real devices: noise and finite sampling. In this section, we discuss the robustness of the multi-class SWAP-Test classifier to noise. We also describe the relationship between the number of label states that can be accurately distinguished on a single qubit as well as how this number is affected by noise.

\subsubsection{Robustness to Noise}
Depolarising noise is a simple yet popular noise model for quantum systems \cite{nielsen2002quantum}. This model, given by a quantum channel 

\begin{equation}
\label{depolarising_error}
	\mathcal{E}_d(\rho) = \frac{pI}{2} + (1-p)\rho,
\end{equation}

\noindent describes the loss of information about a quantum system with some probability $0\le \ p \ \le 1$. The depolarisation parameter $p$ corresponds to the `degree' of depolarising noise where $p=0$ indicates no depolarising noise and $p=1$ indicates a complete loss of information about the system. 

When analysing the effect of depolarising noise on the outcome of the classifier, we consider the same conditions used in \cite{park2020theory} to demonstrate the robustness of the binary SWAP-Test classifier to depolarising noise. We consider the single qubit depolarising channel in equation (\ref{depolarising_error}) but in Kraus form \cite{kraus1971general}, 

\begin{equation}
	\mathcal{E}_d(\rho) = \sum_{k=1}^{4} E_k^{l}\ \rho\ \big( E_k^{l}\big)^\dagger,
\end{equation}

\noindent where the set of Kraus operators are 

\begin{equation}
	E = \bigg\{ \sqrt{1-\frac{3p}{4} I}, \sqrt{\frac{p}{4}} \sigma_x , \sqrt{\frac{p}{4}} \sigma_y, \sqrt{\frac{p}{4}} \sigma_z\bigg\},
\end{equation}

\noindent and the superscript $l$ indicates that the Kraus operator $E_k$ acts only on the label qubit. The Kraus operators also satisfy $\sum_{k=1}^{4} \big( E_k^{l}\big)^\dagger E_k^{l} = \mathbb{I}_2$.

We then evaluate the effect of this channel acting only on the label qubit right before the measurement, that is, on label qubit of the states $\rho_{fx}$, $\rho_{fy}$ and $\rho_{fx}$. The effect of the channel on $\rho_{fx}$ is

\begin{align}
		\mathrm{Tr} \big(\sigma_{z}^{(l)} \mathcal{E}_{d}(\rho_{fx})\big)&=\mathrm{Tr} \big(\mathcal{E}_{d}(\rho_{fx})\big) \sigma_{z}^{(l)})\nonumber\\
		&= \mathrm{Tr} \bigg(\sum_{k=1}^4 E_k^{l} \rho_{fx}\big( E_{k}^{l}\big)^\dagger \sigma_{z}^{(l)}\bigg )\nonumber\\
		&= \sum_{k=1}^4 \mathrm{Tr} \bigg( \rho_{fx} E_k^{l} \sigma_{z}^{(l)} \big( E_{k}^{l}\big)^\dagger \bigg )\\
		&= (1-p)\mathrm{Tr} \bigg(\sigma_{z}^{(l)}\rho_{fx}\bigg)\nonumber
\end{align}

\noindent and, similarly, the effect of the channel on $\rho_{fy}$ and $\rho_{fz}$ is

\begin{align}
	\mathrm{Tr} \big(\sigma_{z}^{(l)}\ \mathcal{E}_d(\rho_{fy})\big) &= (1-p)\mathrm{Tr} \big(\ \sigma_{z}^{(l)}\rho_{fy}\big) \nonumber\\
	\mathrm{Tr} \big(\sigma_{z}^{(l)}\ \mathcal{E}_d(\rho_{fz})\big) &= (1-p)\mathrm{Tr} \big(\ \sigma_{z}^{(l)} \rho_{fz}\big)
\end{align}

\noindent These results can be used to construct a vector

\begin{equation}
	(1-p)\begin{pmatrix}
	\mathrm{Tr} \big( \sigma_z^{(l)} \rho_{fx} \big)\\
	\mathrm{Tr} \big( \sigma_z^{(l)} \rho_{fy} \big)\\
	\mathrm{Tr} \big( \sigma_z^{(l)} \rho_{fz} \big)
	\end{pmatrix} =(1-p) \mathbf{y}_{pred}.
\end{equation}

\begin{figure}[h]
\centering
	\includegraphics[scale=0.4]{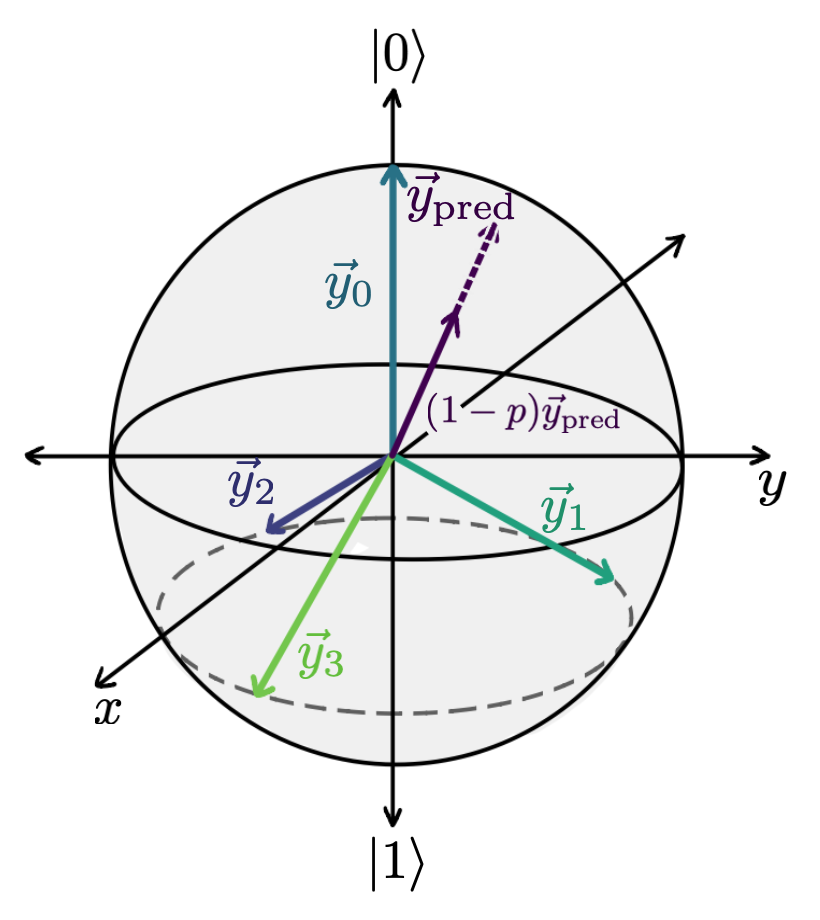}
	\caption{\textbf{An illustrative example of how the outlined conditions of depolarising noise would affect the predicted vector}. The predicted vector will be scaled by a factor of $(1-p)$. This has no impact on the angle between the predicted vector and the label vectors so this will have no impact on the classification outcome. \emph{Note:} this example is not related to any concrete classification problem.}
	\label{scaled_predicted_vectors}
\end{figure}

\noindent This shows that, under the depolarising noise conditions considered, the predicted vector is only scaled by a factor of $(1-p)$. An example of the effect of scaling the predicted is demonstrated in Figure \ref{scaled_predicted_vectors}.

If we rewrite the assignment function in equation (\ref{assignment_function}) as

	\begin{align}
\label{angle_assignment_function}
		\tilde{y} & = \mathrm{max}_{y_i}\{ |\mathbf{y}_i||\mathbf{y}_{pred}|\mathrm{cos}\lambda_i \}\nonumber\\
		& = \mathrm{max}_{y_i}\{ \mathrm{cos}\lambda_i \},
	\end{align}

\noindent we see that the assignment function depends only on the angle $\lambda_i$ between the $i^{th}$ label vector $\mathbf{y_i}$ and the predicted vector $\mathbf{y}_{pred}$. It then becomes apparent that scaling the predicted vector by a factor of $(1-p)$ has no effect on the assignment function. In this way, the outcome of the classifier is unaffected by these conditions of depolarising noise.\\

\subsubsection{Number of Label States}
Our next consideration involves the number of label states that can be stored on a single qubit. Due to the design of the classifier, this number depends on our ability to accurately measure the predicted vector. The number of predicted vectors we can distinguish on the Bloch sphere will be the number of label states that we should store on the single label qubit.

In order to determine the number of predicted vectors that we can distinguish, we first estimate the standard error $\Delta$ in the required measurements. For brevity, we make the following assignments: $x =\mathrm{Tr} ( \sigma_z^{(l)} \rho_{fx} )$, $y=\mathrm{Tr} ( \sigma_z^{(l)} \rho_{fy} )$ and $z=\mathrm{Tr} ( \sigma_z^{(l)} \rho_{fz} )$. Then, the standard error for $s\in\{x,y,z\}$ is

\begin{equation}
		\Delta s = \sqrt{\frac{4 p_s(1-p_s)}{R}} 
\end{equation}

\noindent where $p_s= \frac{1}{2} \big( \mathrm{Tr} \big( \sigma_z^{(l)} \rho_{fs} \big)+1 \big)$ and $R$ is the number of repetitions.  

Using the standard error propagation formula, we estimate $\Delta \theta$ and $\Delta \phi$. This allows us to estimate the area of the ellipsoid that represents the uncertainty in the measurement of the predicted vector. An illustration of this ellipsoid is in Figure \ref{ellipsoids}. By dividing the surface area of the Bloch sphere by the area of this ellipsoid, we find the number of label states ($N_s$) that can be stored:

\begin{figure*}
\centering
	\includegraphics[scale=0.4]{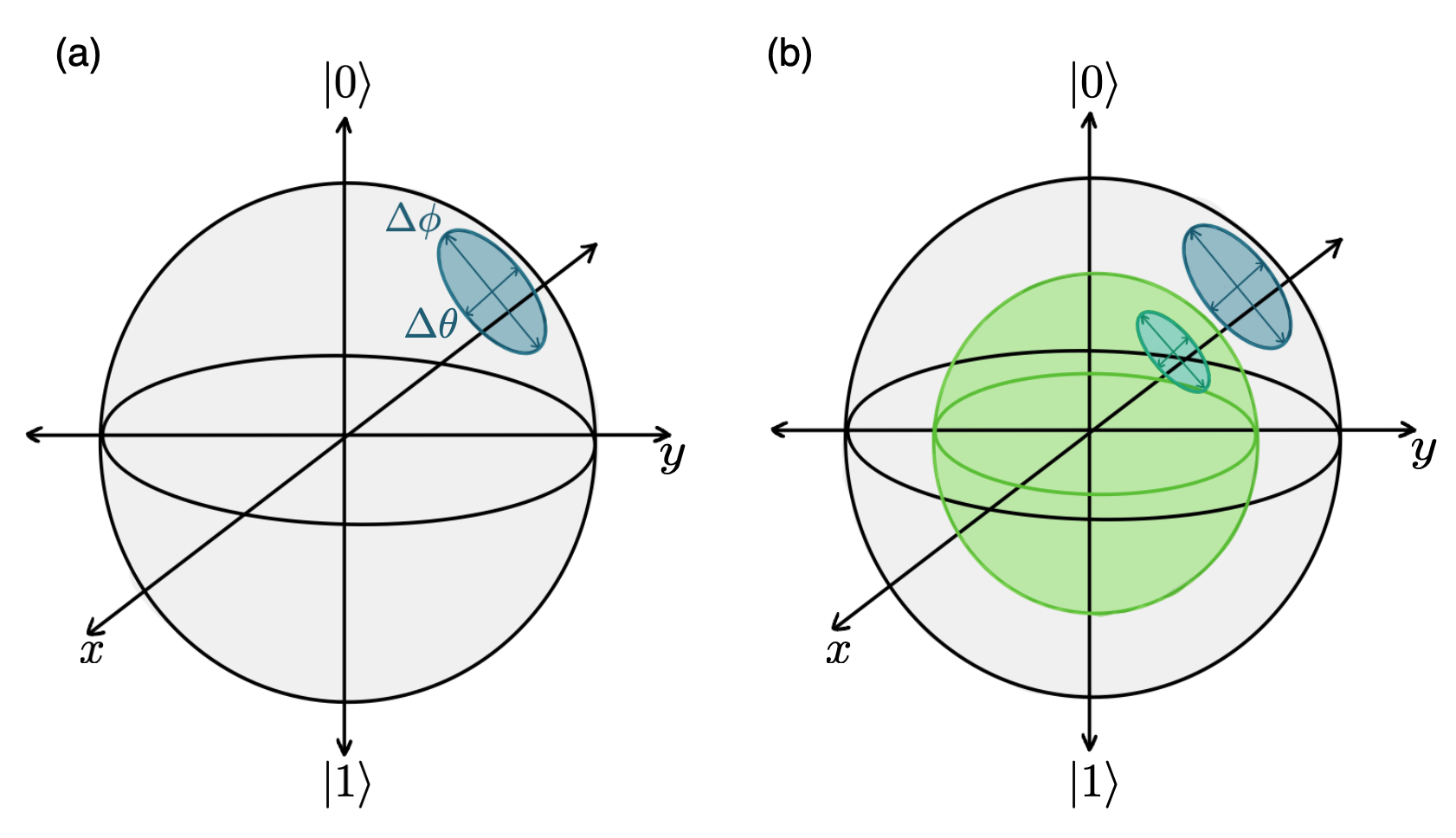}
	\caption{\textbf{Ellipsoids representing the standard error in the measurement of a predicted vector}. (a) shows the ellipsoid on the Bloch sphere while (b) shows the ellipsoid on a shrunken Bloch sphere which can be expected under certain depolarising noise conditions. In each case, the number of label states that should be stored on the label qubit can be calculated by dividing the area of the sphere by the area of the ellipsoid.}
	\label{ellipsoids}
\end{figure*}

\begin{align}
	\label{number_of_classes_no_noise}
		N_s & = \frac{{\mathrm{Area\ of\ Bloch\ Sphere}}}{\mathrm{Area\ of\ Ellipsoid}}\nonumber\\
		& = \frac{4 \pi ||\vec{r}||^2}{\pi \Delta \theta \Delta \phi}
	\end{align}

\noindent where $\vec{r} = (x,y,z)$.  

After making the necessary substitutions, we find that

\begin{equation}
	N_s = \mathcal{O}(R).
\end{equation}

\noindent This means that the number of label states that can be stored grows linearly with the number of repetitions of the required measurements. 

If we consider the effect of depolarising noise on this number, then the number of label states that can be stores under these conditions ($\tilde{N}_s$) is:

\begin{align}
\label{number_of_classes_with_noise}
	\tilde{N}_s & = \frac{4 \pi ||(1-p)\vec{r}||^2}{\mathrm{Area\ of\ Ellipsoid}}\nonumber\\
	& = \frac{4 \pi (1-p)^2||\vec{r}||^2}{A(p)},
\end{align}

\noindent where the $\mathrm{Area\ of\ Ellipsoid}$ becomes a function $A(p)$ that is now also dependent on the depolarising parameter $p$. 

A Taylor series expansion of $A(p)$ around the origin ($p=0$), up to second order, reveals: 

\begin{equation}
	A(p) = A_0+A_1p+A_2p^2+\mathcal{O}(p^3),
\end{equation}

\noindent where $A_1 = \frac{\partial{A(p)}}{\partial{p}}|_{p=0}$ and $A_1 = \frac{1}{2}\frac{\partial^2{A(p)}}{\partial{p}^2}|_{p=0}$.

By substituting this expansion in equation (\ref{number_of_classes_with_noise}), we obtain an expression for the number of label states that accounts for depolarising noise:

\begin{align}
 		\tilde{N}_s & = \frac{4 \pi (1-p)^2||\vec{r}||^2}{A(p)}\nonumber\\
 		& \simeq  \frac{4 \pi (1-p)^2||\vec{r}||^2}{\frac{1}{A_0}(1+\frac{A_1}{A_0}p+\frac{A_2}{A_0}p^2)}\nonumber\\
 		& \simeq \frac{4\pi||\vec{r}||^2}{A_0}(1-2p+p^2)[1-\frac{A_1}{A_0}p-(\frac{A_2}{A_0}-\frac{A_1^2}{A_0^2}p^2)]\nonumber\\
 		& \simeq N_s\bigg[1-(\frac{A_1}{A_0}+2)p-(\frac{A_2}{A_0}-\frac{A_1^2}{A_0^2}-2\frac{A_1}{A_0}-1)p^2\bigg].
\end{align}

A numerical search for the maximum values of $(\frac{A_1}{A_0}+2)$ and $(\frac{A_2}{A_0}-\frac{A_1^2}{A_0^2}-2\frac{A_1}{A_0}-1)$ with $-1\le x, y, z \le1$ and $\sqrt{x^2+y^2+z^2}\le1$ yields

\begin{align}
		(\frac{A_1}{A_0}+2) &\le 5\\
		(\frac{A_2}{A_0}-\frac{A_1^2}{A_0^2}-2\frac{A_1}{A_0}-1) &\le -6.
\end{align}

\noindent This reveals that in the worst case:

\begin{equation}
\label{number_of_classes_with_noise_worst_case}
		\tilde{N}_s \sim N_s(1-5p+6p^2).
\end{equation}

\noindent These conditions of depolarising noise decrease the number of label states by at most $(1-5p+6p^2)$. The depolarising parameter that we expect in real experiments is $p<<0.1$. According to equation (\ref{number_of_classes_with_noise_worst_case}), this will have a negligible effect on the number of label states that can be stored. 
 
 \section{Methods}
 
 We demonstrate the effectiveness of the multi-class SWAP-Test classifier by applying it to several diverse multi-class classification problems.
 
\begin{figure}
\centering
	\includegraphics[scale=0.35]{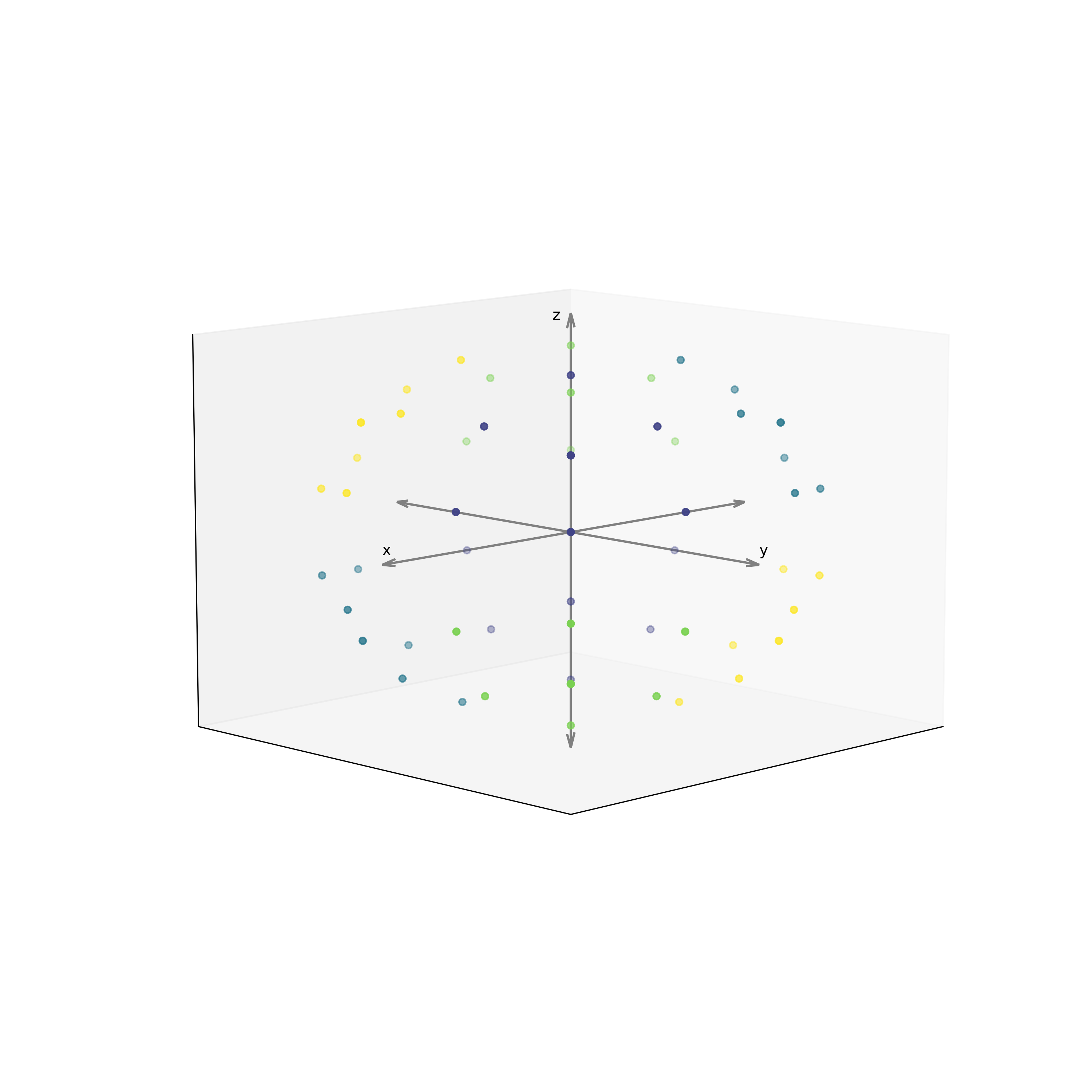}
	\caption{\textbf{The $\texttt{4-XOR}$ dataset (3 features, 4 classes and 64 points)}. The dataset was generated such that data points from the same class would lie directly opposite each other on a sphere so that the kernel in equation (\ref{linear_kernel}) would be most effective.} 
	\label{small_xor_3d}
\end{figure}

 For the numerical simulations, we apply the classifier to a 3-dimensional generated dataset with 4 classes; 16 data points in each class. This dataset, shown in Figure \ref{small_xor_3d}, will be denoted by $\texttt{4-XOR}$ since it is inspired by the 2-dimensional XOR dataset. Like the 2-dimensional XOR dataset, $\texttt{4-XOR}$ is not linearly separable. 
 
\begin{figure*}
\centering
	\includegraphics[scale=0.4]{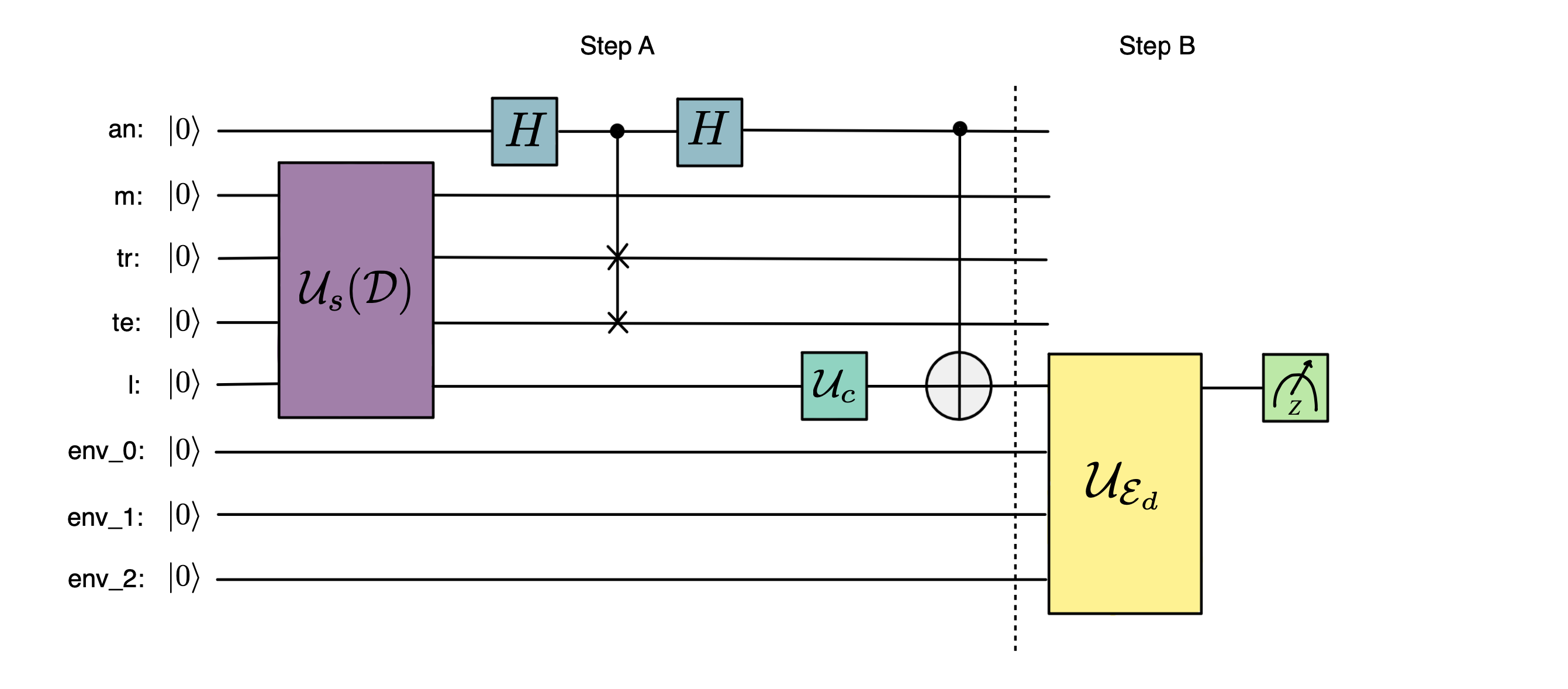}
	\caption{\textbf{The circuit required to simulate the multi-class SWAP-Test classifier with depolarising noise}. Step A involves the state preparation, modified SWAP-Test and change of basis. Step B involves applying $\mathcal{U}_{\mathcal{E}d}$ to simulate the depolarising noise.}
	\label{circuit_with_noise_diagram}
\end{figure*}

\begin{figure*}
\centering
	\includegraphics[scale=0.4]{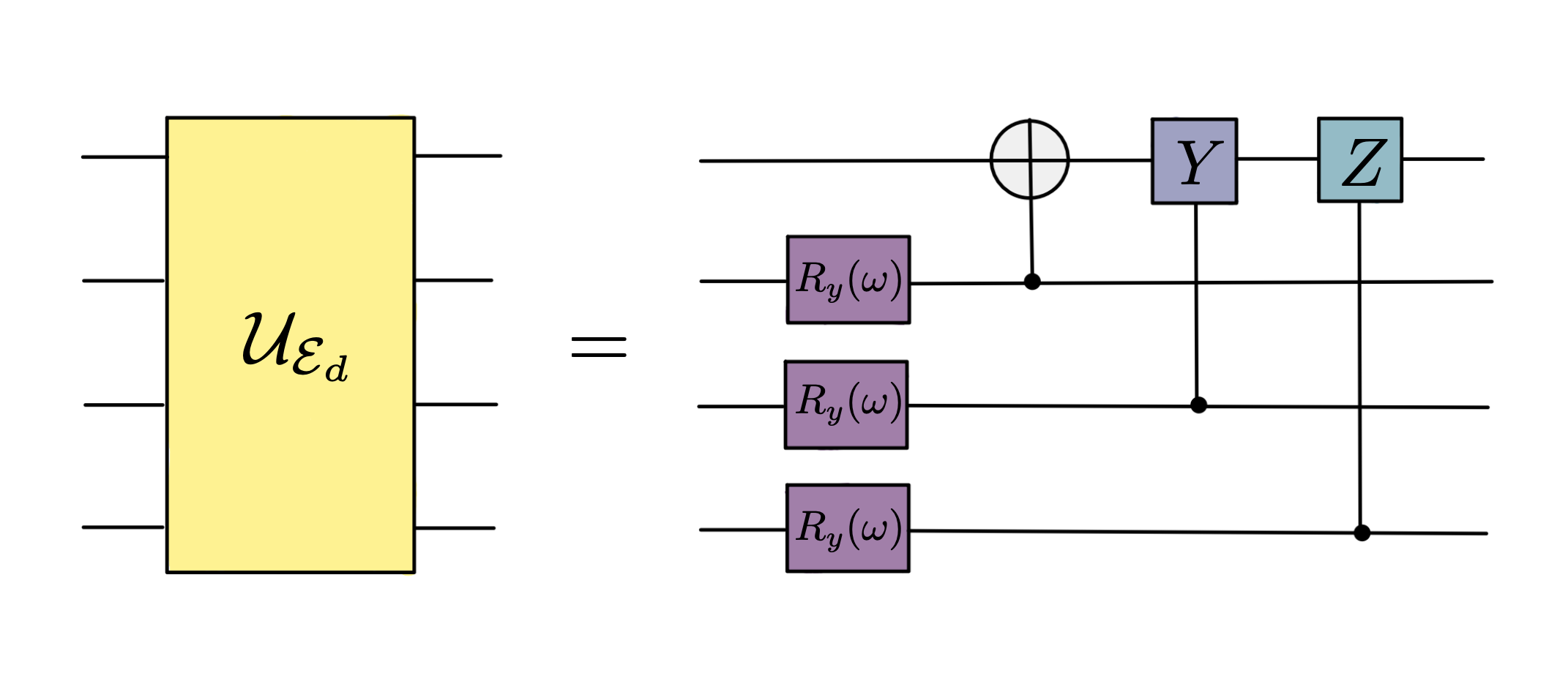}
	\caption{\textbf{A more detailed look at the operator $\mathcal{U}_{\mathcal{E}d}$ that simulates the depolarising noise in the circuits}. Here, $R_y$ is a Y-rotation with $\omega = \frac{1}{2} \mathrm{arccos}(1-2p)$.}
	\label{expanded_depolarising_circuit}
\end{figure*}

 To evaluate the performance of the multi-class SWAP-Test classifier on this dataset, we perform leave-one-out cross validation. For each data point, we numerically simulate the circuits that construct the predicted vector using Qiskit. The predicted vector is then used in the assignment function given in equation (\ref{assignment_function}) to classify the data point. We first numerically simulate the circuits ideally, with no noise or finite sampling. We then numerically simulate the circuits with finite sampling and under the depolarising noise conditions outlined in Section 2.2.1. Figure \ref{circuit_with_noise_diagram} depicts the circuits that are simulated with depolarising noise, with Figure \ref{expanded_depolarising_circuit} providing a more details on the simulation of the depolarising noise \cite{garcia2020ibm}.  
 
 When encoding one of the test or training data, we first normalise the datum so that $\norm{\mathbf{x}}^2 = \sum_i |x_i|^2 =1$. Then, we encode each component of the normalised datum in a different computational basis state of the appropriate register: 
 
 \begin{equation}
 	\mathbf{x} \to \sum_{i=1}^{2^n} x_i \ket{i}.
 \end{equation}
 
 This strategy is commonly known as amplitude encoding and it gives rise to the following quantum kernel:
 
\begin{equation}
\label{linear_kernel}
	k(\mathbf{x},\mathbf{z}) = |\braket{\mathbf{x}}{\mathbf{z}}|^2,
\end{equation}

which is simply the absolute square of the linear kernel \cite{schuld2019quantum}. It should also be noted that uniform training weights are used, that is: $\sum_{m=1}^{M} w_m = 1,\ w_m=1/M \ \forall \ M$. 
 
\begin{figure}[h]
\centering
	\includegraphics[scale=0.3]{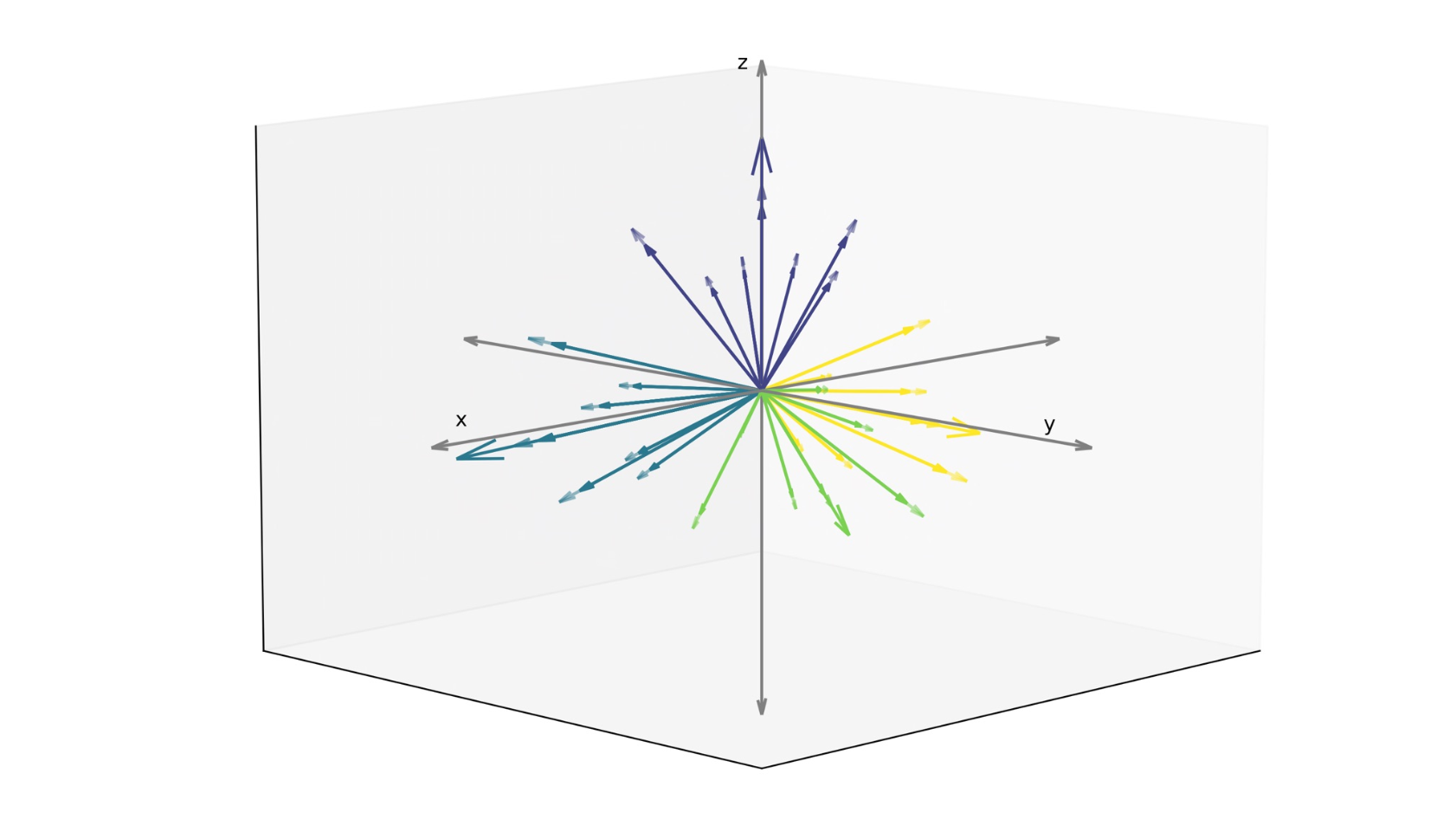}
	\caption{\textbf{The predicted vectors produced through the classification of the $\texttt{4-XOR}$ dataset}. Here, the darker vectors are the predicted vectors that have been evaluated with depolarising noise ($p=0.1$) while the lighter vectors are the predicted vectors evaluated without depolarising noise. We can see that the predicted vectors are only scaled down but the angles between the vectors does not change.}
	\label{noisy_and_noiseless_predicted_vector}
\end{figure}

 Our simulations show that high accuracies (100\%) are obtained by the classifier when executed under ideal conditions and realistic conditions with depolarising noise and finite sampling. The results of these experiments are shown in Table \ref{small_xor_results}. Upon analysis of the predicted vectors obtained from simulations with depolarising noise, the result from Section 2.2.2 is confirmed. We see that the obtained predicted vectors are only scaled under the depolarising noise conditions and that this has no effect on the classification process. This is illustrated in Figure \ref{noisy_and_noiseless_predicted_vector}. This is also illustrated in Table \ref{small_xor_results} where we see that the average norms of the predicted vectors decreases by a factor of $(1-p)$.
 
\begin{table*}[h]
\centering
\caption{The accuracies obtained by the multi-class SWAP-Test classifier when applied to the generated $\texttt{4-XOR}$ dataset. The accuracies presented are from numerical simulations with finite sampling and depolarising noise. The average norms of the predicted vectors are from numerical simulations with just depolarising noise, to illustrate the result from Section 2.2.1}
\begin{tabular}{ccccc}
\hline
Depolarisation Rate ($p$) & & Accuracy (\%) & & Av. Norm of Predicted Vector \\
\hline
0            &       & 100    &  & 0.1356                             \\
0.02          &      & 100   &   & 0.1331                           \\
0.04          &      & 100   &   & 0.1302                            \\
0.06        &        & 100   &   & 0.1275                            \\
0.08        &        & 100    &  & 0.1248                           \\
0.1         &        & 100    &  & 0.1223 \\
\hline
\end{tabular}
\label{small_xor_results}
\end{table*}
 
 We further evaluate the performance of the multi-class classifier on several other datasets using 5-fold cross validation. These datasets include three generated XOR datasets as well as the Iris, Wine and Digits datasets provided by scikit-learn. It should be noted that the Wine and Digits dataset were artificially balanced. For the Wine dataset, this was done by uniformly sampling 48 data points from each class in each dataset. For the Digits dataset, 174 points from each class were uniformly sampled. 
 
 For each test point in these datasets, we evaluate the predicted vector classically using equation (\ref{y_pred_eq}). To do this, we evaluate each $\alpha_i = \sum_{m|\mathrm{y}_m=i} \mathrm{w}_m | \langle \mathbf{\tilde{x}} |\mathbf{x}_m\rangle|^2$ directly as $\alpha_i = \sum_{m|\mathrm{y}_m=i} \mathrm{w}_m k(\mathbf{\tilde{x}},\mathbf{x}_m)$. This is only possible because the methods that we would use to encode the test and training data in quantum states give rise to kernels $k(\mathbf{\tilde{x}},\mathbf{x}_m)$ that can be evaluated classically. In some cases, we use the kernel $k(\mathbf{x},\mathbf{z}) = |\braket{\mathbf{x}}{\mathbf{z}}|^2$ which would arise from amplitude encoding. In other cases, we use the kernel $k(\mathbf{x},\mathbf{z})= \prod_{k=1}^{n} |\mathrm{cos(x_k-z_k)}|^2$ which would arise from angle encoding.  Once the predicted vector is constructed, it is then used in the assignment function given in equation (\ref{assignment_function}) to classify the test point. The results of these experiments can be seen in Table \ref{other_datasets_results}. It can be seen that the accuracies are high, with accuracies above 90$\%$ being obtained for each dataset. 

\begin{table*}[htb]
\centering
\caption{The accuracies obtained by the multi-class SWAP-Test classifier when applied to various datasets.}
\begin{tabular}{cccccc}
\hline
Dataset & \# Classes & \# Features & \# Points & Encoding & Accuracy (\%) \\ \hline
XOR     & 2               & 2              & 100           & Amplitude & 100           \\
       & 4               & 3              & 200           & Amplitude & 99            \\
       & 8               & 4              & 400           & Amplitude & 99            \\ \hline
Iris    & 3               & 4              & 150           & Angle & 95            \\ \hline
Wine    & 3               & 13             & 144           & Angle & 92            \\ \hline
Digits  & 10              & 64             & 1740          & Angle & 92            \\ \hline
\end{tabular}
\label{other_datasets_results}
\end{table*}

\section{Conclusion}
We presented a multi-class quantum classifier comprising of a set of quantum circuits and an assignment function that is evaluated classically. The quantum circuits estimate a weighted sum of kernel values between the test and training data and the assignment function allows us to meaningfully interpret this sum. Analytically and through experiments, the multi-class SWAP-Test classifier has been shown to be powerful and robust to noise. It has also been shown that the classifier can handle $\mathcal{O}(R)$ classes, where $R$ is the number of executions of the required circuits. In our numerical experiment, a standard encoding strategy has been employed and has proven effective on the generated dataset. It should be noted that for future work, encoding strategies that give rise to kernels that are classically intractable to simulate \cite{havlivcek2019supervised} can be chosen. Like the binary SWAP-Test classifier, this classifier can also raise the evaluated kernels to the power of $n$ at the cost of $n$ copies of the test and training data. Our numerical experiments also utilised uniform training weights but future work could develop a strategy that utilises trainable or non-uniform training weights. Also of interest for future work would be the use of real quantum devices to solve multi-class classification problems with this classifier. 

\section*{Acknowledgements}
This work is based upon research supported by
the National Research Foundation of the Republic of
South Africa. Support from the NICIS (National Integrated Cyber Infrastructure System) e-research grant
QICSA is kindly acknowledged. Support from the CSIR DSI-Interbursary Support (IBS) Programme is gratefully acknowledged. Support from the Center of Artificial Intelligence Research is appreciated. We acknowledge the use of IBM Quantum for the use of their simulators. The views expressed are those of the authors and do not reflect the official policy or position of IBM or the IBM Quantum team. We would like to thank Mr A. W. Pillay and Mr I. J. David for their assistance in proof reading the manuscript. 

\bibliographystyle{unsrt}
\bibliography{bibliography}

\end{document}